\documentclass[prd,
preprint,aps,amsfonts,amssymb,nofootinbib,tightenlines]{revtex4}

\usepackage{amscd}
\usepackage{amsmath}
\usepackage{bm}
\usepackage[dvips]{color}                 
\usepackage{graphicx}
\usepackage{eufrak}
\usepackage{eucal}

\newcommand{\beq}{\begin{equation}}
\newcommand{\eeq}{\end{equation}}
\newcommand{\bea}{\begin{eqnarray}}
\newcommand{\eea}{\end{eqnarray}}

\newcommand{\nn}{\nonumber}
\newcommand{\benn}{\begin{displaymath}}
\newcommand{\eenn}{\end{displaymath}}
\newcommand{\ket}[1]{| #1 \rangle}                     
\newcommand{\bra}[1]{\langle #1 \, |}                  
 
\newcommand{\refeq}[1]{(\ref{eq:#1})} 
\newcommand{\Eq}{Eq.~\refeq}
\newcommand{\eqn}[1]{\label{eq:#1}}

\def\ma{{M_A}}

\def\mua{\mu_A}

\def\ea{\epsilon^A}
\def\eb{\epsilon^B}
\def\eplus{ {\epsilon^+} }

\def\psia{\psi_A}
\def\psib{\psi_B}
\def\[{\left[}
\def\]{\right]}

\begin{document}
\preprint{LBNL-53196}

\title{ Phase separation in asymmetrical fermion superfluids}

\author{Paulo F. Bedaque\footnote{Email: {\tt pfbedaque@lbl.gov}}, 
Heron Caldas\footnote{Email: {\tt hcaldas@lbl.gov}. 
Permanent address: Universidade Federal de Sao Joao del Rey, S\~{a}o Jo\~{a}o del Rei, 36300-000, MG, Brazil.} and
Gautam Rupak\footnote{Email: {\tt 
grupak@lbl.gov}} }
\affiliation{Lawrence-Berkeley Laboratory, Berkeley, CA 94720, U.S.A.}

\begin{abstract}
Motivated by recent developments on cold atom traps and high density QCD
we consider 
fermionic systems composed of two particle species with  different densities. 
We argue that a mixed phase composed of normal and superfluid components is the energetically favored ground state. 
We suggest how this phase separation can be used as a probe of fermion superfluidity in 
atomic traps.
\end{abstract}
\maketitle
\bigskip
The recent interest in two physical systems have revived the study of 
asymmetric
fermionic systems, that is, systems with unequal  number density (or chemical
potential) for the different species.  In high density strongly interacting
systems, as it may be found at the core of ``neutron'' stars, the different
quark flavors have different chemical potentials on account of their different
masses and 
charges~\cite{krishna_review,Rischke:2003mt,Schafer:2003vz,Alford:2001dt}. 
Atomic traps can also provide an
example of similar systems if a bias can be introduced when the trap is filled
in order to have a larger number of one of the atom species (or hyperfine state
of the same atom)~\cite{pethick_book}. In the symmetrical situation the low
energy properties of these systems are dominated by Cooper pair formation. 
Since
it involves the attraction between two fermions with equal and opposite momenta
at the their Fermi surface one could imagine that as the Fermi surfaces move
apart with increasing asymmetry, the pairing would become weaker and the gap
smaller, until superfluidity disappears. What actually happens, the formation 
of
an inhomogenious mixed phase state, is however more interesting and results 
from the competition between states with different particle distribution, 
both in momentum and real space.

Our discussion is valid in a wider class of models but, for definiteness, let 
us consider a  non-relativistic dilute gas made out of two particle species 
$A$ and
$B$ with chemical potentials $\mu_A$, $\mu_B$ and masses $M_A$, $M_B$
respectively. 
At low
densities the details of the potential are not probed and their interaction is
well described by the pairing Hamiltonian 
\begin{align}
\mathcal{H} - \sum_{i=A,B}\mu_i \mathcal{N}_i 
= \int \frac{d^3k}{(2\pi)^3}\sum_{i=A,B}\epsilon^i_k \psi_i^\dagger(k) 
\psi_i(k)
+ g \int \frac{d^3k}{(2\pi)^3} \frac{d^3p}{(2\pi)^3}\psia^\dagger(p) 
\psib^\dagger(-p) \psib(-k)\psia(k),
\end{align}
where $\psia^\dagger, \psia$ are creation and annihilation operators for the
 $A$ particles and $\ea_k$ is their dispersion relation, that we 
take to be
$\ea_k= k^2/2\ma - \mua$ (and similarly for $B$).
In the mean field approximation, adequate for the low densities considered here, the Hamiltonian can be approximated by 
\begin{align}\eqn{omegadiagonal}
\mathcal{H}- \sum_{i=A,B}\mu_i \mathcal{N}_i 
&= -\frac{|\Delta|^2}{g}+\int \frac{d^3k}{(2\pi)^3} 
\sum_{i=A,B}\epsilon^i_k  \psi_i^\dagger(k)\psi_i(k) 
-\Delta^*\psib(-k)\psia(k) -\Delta\psia^\dagger(k)\psib^\dagger(-k)
\nn\\
&= -\frac{\Delta^2}{g}+\int \frac{d^3k}{(2\pi)^3}(\eb_k-E_k^\beta)+
\int \frac{d^3k}{(2\pi)^3}
\[E^\alpha_k \psi_\alpha^\dagger(k)  \psi_\alpha(k)
+E^\beta_k \psi_\beta^\dagger(k)\psi_\beta(k)\],
\end{align}
where $\Delta = 
-g \int d^3k/(2\pi)^3\langle \psib(-k)\psia(k)\rangle=\Delta^\ast$, $\ E^{\alpha,\beta}_k = \pm\epsilon^-_k+\sqrt{\epsilon^{+ 2}_k+\Delta^2}$,  $\epsilon^{\pm} = (\ea_k \pm \eb_k)/2$ 
 and the fields $\psi_\alpha,\psi_\beta$ are defined by
\beq
\begin{pmatrix}
\psi_\alpha(k) \\
\psi_\beta^\dagger(-k)
\end{pmatrix}
=
\begin{pmatrix}
u_k & -v_k\\
v_k & u_k
\end{pmatrix}
 \begin{pmatrix}
\psia(k) \\
\psib^\dagger(-k)
\end{pmatrix}
, \eeq with
\beq
\begin{matrix} u_k^2\\
                v_k^2
                \end{matrix}=
 \frac{1}{2}\left( 1\pm\frac{\eplus_k}{\sqrt{\eplus_k^2 +\Delta^2}} \right)  .
\eeq 
It is straightforward to minimize the diagonalized Hamiltonian shown in 
\Eq{omegadiagonal}. 
One simply fills the modes with negative $E^{\alpha,\beta}_k$ and leave the remaining modes empty. More precisely, the ground state $\ket{\Psi}$ satisfies
\begin{align}
\psi_{\alpha,\beta}(k) \ket{\Psi}  &= 0 \quad \text{if}
\quad E^{\alpha,\beta}_k >0,\nn\\
\psi_{\alpha,\beta}^\dagger(k) \ket{\Psi} &= 0 \quad \text{if}
\quad E^{\alpha,\beta}_k <0.
\end{align}

In terms of the original particles $A$ and $B$ and the vacuum state 
$\ket{0}$,
the state above corresponds to having a BCS-like state 
$[u_k+v_k \psia^\dagger(k)\psib^\dagger(-k)]\ket{0}$ in the modes $k$ where
$E^{\alpha,\beta}_k >0$, but a state filled with  
particle $B$ ($A$) 
in the modes where  $E^{\beta}_k <0$ ($E^{\alpha}_k <0$). 
The thermodynamic potential of this state is
\begin{align}\eqn{omega_mean}
\Omega =\bra{\Psi}{\mathcal{H} 
- \sum_{i=A,B}\mu_i \mathcal{N}_i }\ket{\Psi}
= -\frac{M\Delta^2}{2\pi a} + \int \frac{d^3k}{(2\pi)^3}\left[ \theta(-E^\alpha_k)E^\alpha_k + \theta(-E^\beta_k)E^\beta_k +\epsilon^B_k - E^\beta_k\right],
\end{align}
where $a$ is the scattering length between particle $A$ and $B$.
It is related to the coupling constant (in dimensional regularization) by
$1/g = M/(2\pi a)$ with  $M=M_A M_B/(M_A+M_B)$, the reduced mass.

\begin{figure}[t]
\includegraphics[height=2in]{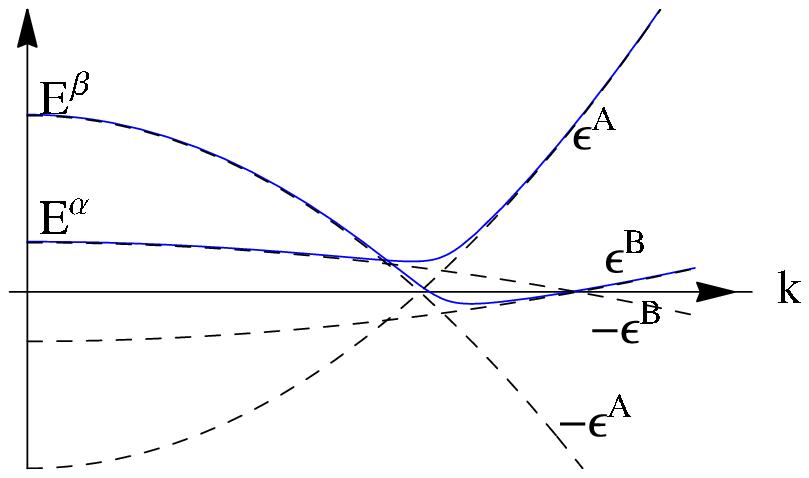}
\caption{\label{dispersion}\textit{  Dispersion relation for the quasi-particles
$\alpha$ and $\beta$ showing a region where $E^\beta_k$ is negative for
$M_B= 7 M_A$, $p_B=1.4\, p_A$, and  $\Delta=0.02 \mu_A<
(p_B^2-p_A^2)/(4\sqrt{M_A M_B})$. Solid curve correspond to $E_k^\beta$ (top)
and $E_k^\alpha$ (bottom).    
} } 
\end{figure}

Let us now consider the case $M_B>M_A$ , $p_B>p_A$, where $p_i$ is 
the Fermi momentum  defined
by $p_i=\sqrt{2 M_i\mu_i}$. Fig.~\ref{dispersion} shows that, for some values of
$\Delta$, $E^\beta_k$ may be negative for momenta $k_1\le k\le k_2$ where
\begin{align}
k_{1,2}^2 = \frac{p_A^2+p_B^2}{2}\pm \frac{1}{2}\sqrt{(p_B^2-p_A^2)^2-16
M_AM_B\Delta^2}\ ,
\end{align}while $E^\alpha_k$ is always positive.

We now discuss separately the cases where either the chemical potentials or the densities of each species are kept fixed.

\subsubsection{Fixed chemical potentials}

\begin{figure}[t]
\includegraphics[height=1.6in]{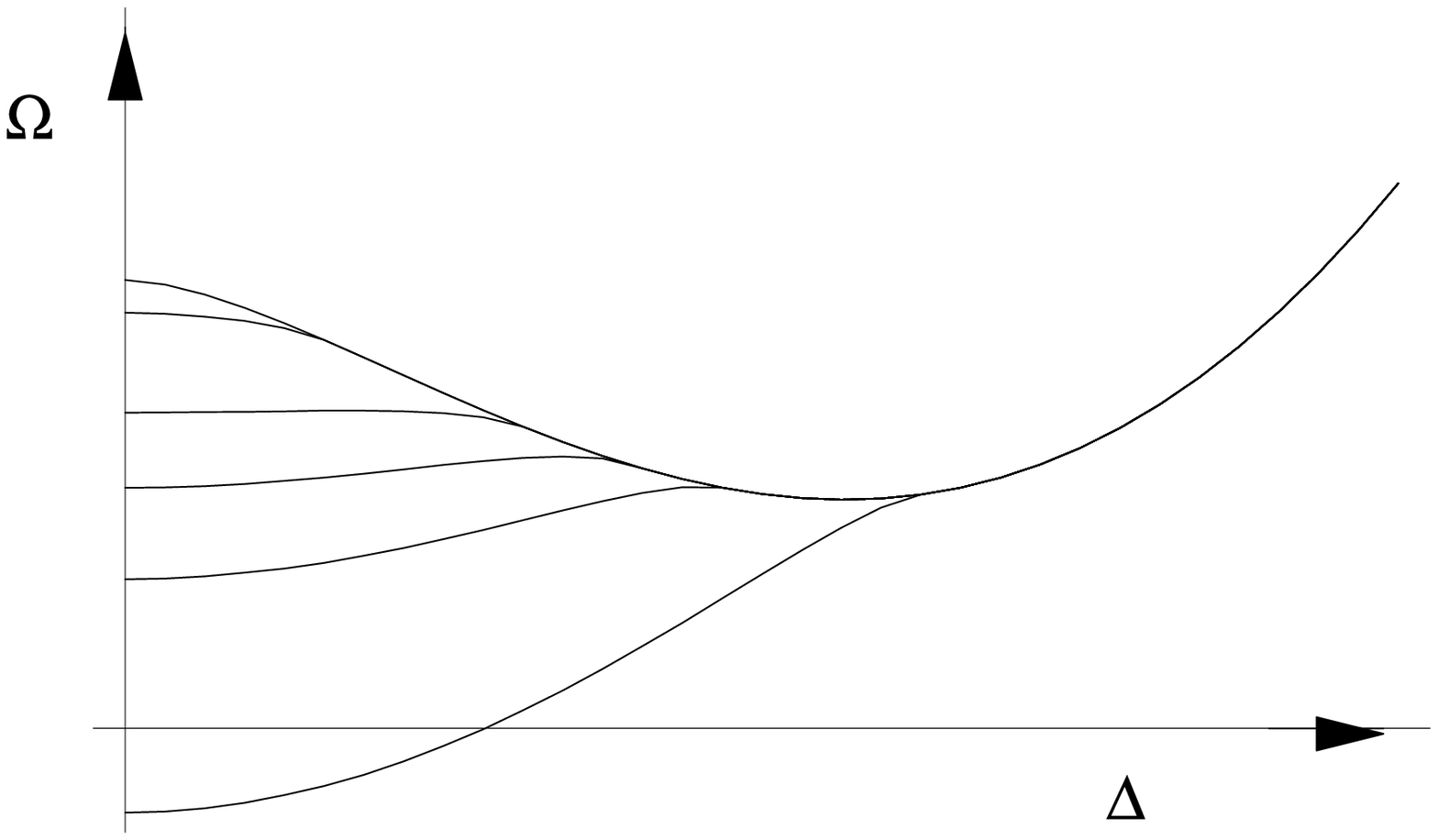}
\caption{\label{omega}\textit{Thermodynamic potential for different values of
$p_B$ and $p_A$ (constant $p_0$). The top curve corresponds to $p_A=p_B$ and 
the
lower curves correspond to increasing values of $|p_B^2-p_A^2|$.} } 
\end{figure}

In Fig.~\ref{omega} we show the thermodynamic potential as a function of
$\Delta$ for different values of $p_A$ and $p_B$,  keeping the combination
$p_0^2/M=p_A^2/M_A + p_B^2/M_B$ fixed, computed from a numerical evaluation 
of \Eq{omega_mean}. An analytical expression valid for 
$\Delta \ll \mu_{A,B}$ is
available but it is not very enlightening.

 For large enough $\Delta$, $k_{1,2}^2$ are not real, $E^{\alpha,\beta}_k$ is
always positive and the thermodynamic potential is unchanged from the $p_B=p_A$
case. However, for $\Delta < (p_B^2-p_A^2)/(4 \sqrt{M_A M_B})$, 
$k_{1,2}^2 $ are
positive and the thermodynamic potential can be lowered by filling the states
between $k_1$ and $k_2$ with $\beta$-type quasi-particles. The $\Delta=0$ state,
in particular, has its thermodynamic potential lowered with increasing
$p_B^2-p_A^2$ and at some point becomes smaller than the previous minimum with
$\Delta=\Delta_0$ corresponding to the  BCS phase. The result is that, for fixed
$p_0$ and increasing $p_B^2-p_A^2$, there is a first-order phase transition
between the superfluid  and the normal state, as it has been noticed in
different physics contexts (see, for instance Refs.~\cite{glendenning,
paulo_asym, sedrakian, alford_unlocking}). These results can be understood in
very simple physical terms. Suppose we start in the BCS phase with $\mu_A=\mu_B$ and increase $p_B$. By absorbing one $B$ particle and eliminating one $A$ the system by one hand reduces its thermodynamic potential $\Omega$ due to the $-\mu_A n_A-\mu_B n_B$ term but by the other hand it increases $\Omega$ by destroying two pairs. This is energetically favorable if, and only if, the difference in chemical potentials is large enough (or the gap small enough). Until that point the BCS state with equal number of particles remains the ground state, unchanged despite the variations in chemical potential.
 
 In addition to the stable (or metastable) normal and BCS phases there is, for
some values of the chemical potentials $\mu_A$ and $\mu_B$, 
an unstable phase (referred to here as
``Sarma state") corresponding to a maximum of $\Omega$ as a function of $\Delta$
situated between the BCS minimum $\Delta_0$ and the normal phase at 
$\Delta=0$ (first pointed out in Ref.~\cite{sarma}). The combination of
parameters necessary for the existence of this phase can be found considering
the  gap equation: 
\begin{align}
 0 &= \frac{d\Omega}{d\Delta^2}=-\frac{M}{2\pi a}
-\frac{1}{2}\int \frac{d^3k}{(2\pi)^3}
 \frac{1}{\sqrt{\epsilon^{+2}_k + \Delta_s^2}}
+\frac{1}{2}\int_{k_1}^{k_2} \frac{d^3k}{(2\pi)^3} \frac{1}{\sqrt{\epsilon^{+2}_k + \Delta_s^2}}\nn\\
 &= -\frac{M}{2\pi a}-\frac{1}{2}\int \frac{d^3k}{(2\pi)^3}
 \frac{1}{\sqrt{\epsilon^{+2}_k + \Delta_0^2}},
\end{align}
where the BCS gap is given by
\begin{align}
 \Delta_0 \cong 2\frac{p_0^2}{M}\ e^{-\frac{\pi}{2 p_0 |a|}-2},
\end{align}
in the mean field approximation.
 For small values of the gaps $\Delta_0,\Delta_s \ll \mu_A,\mu_B$ the integrals can be approximated and it is found that~\cite{sarma,breached,yip}
  \begin{align}\eqn{deltas}
 \Delta_s \cong \sqrt{\Delta_0 \left(
\frac{p_B^2-p_A^2}{2\sqrt{M_AM_B}}-\Delta_0  \right)}\ .
 \end{align}
The Sarma state gap is in the range $0\le \Delta_s\le \Delta_0$, where the upper
bound is set by 
the condition for the existence of real values of $k_{1,2}^2$. We then have 
\begin{align}\eqn{conditionforsarma}
0\le\sqrt{\Delta_0 \left(
\frac{p_B^2-p_A^2}{2\sqrt{M_AM_B}}-\Delta_0  \right)}\le \Delta_0 \Rightarrow
\Delta_0 \leq \frac{p_B^2-p_A^2}{2\sqrt{M_AM_B}} \leq 2 \Delta_0.
 \end{align}
  The condition in \Eq{conditionforsarma}, for fixed $p_{A}$ and $p_B$,
 can thus
be seen as determining a window for the values of $a$ supporting the Sarma
state. It does not exist, even as an unstable state, for too small or too large
interactions~\cite{gapless}. Also notice that when $p_B^2-p_A^2$ reaches its {\it largest}
value allowing for a Sarma state ($p_B^2-p_A^2=4\sqrt{M_AM_B}\Delta_0$), $k_1$
equals $k_2$ and the difference in particle densities $n_B-n_A$ approaches
 zero, corresponding to a BCS state. In the other limit, $p_B^2-p_A^2=2\sqrt{M_AM_B}\Delta_0$, the gap $\Delta_s$ vanishes and the Sarma state reduces to the normal phase.
 
 \subsubsection{Fixed particle number}
 The previous discussion regarding the  stability of different phases was made under the assumption that the particle numbers $n_A$ and $n_B$ are allowed to change. We discuss now the situation where they are fixed. 
 In the BCS phase the particle number densities $n_A$ and $n_B$  are the same,
as can be readily seen by taking derivatives of \Eq{omega_mean} in relation to
$\mu_A$ and $\mu_B$, while the particle numbers in the normal phase can be different. 
 The Sarma phase can also accommodate $n_A\neq n_B$ so the question arises: 
what is the ground state of the system when the particle densities are fixed and
different from each other? This question has been revived in  
 Refs.~\cite{internalgap,internalgap_comment,gapless} 
where it was argued that the
Sarma phase (named there ``internal gap"' state) could be the ground state in
the case of fixed particle numbers $n_A$, $n_B$.
 This state would have fascinating properties, being at the same time  a Fermi liquid (with two Fermi surfaces corresponding to $k_1$ and $k_2$) {\it and} a superfluid.

The question can be answered 
by finding the state with the smallest energy (not the thermodynamic potential $\Omega$). We compare here the the normal, Sarma and a mixed inhomogeneous phase composed of bubbles of normal phase in a sea of the BCS phase.
The  energy for the normal and the BCS phase are, for small values of the gap, given by
\begin{align}
E_N(n_A,n_B) &= \frac{(6\pi^2n_A)^{5/3}}{20 \pi^2 M_A}+ \frac{(6\pi^2n_B)^{5/3}}{20 \pi^2 M_B},\nn\\
E_{BCS}(n_A=n_B=n) &= \frac{(6\pi^2 n)^{5/3}}{20\pi^2 M}-\frac{M\Delta_0^2(n)}{2\pi^2}(6\pi^2 n)^{1/3}.
\end{align} 
A similar expression can be derived for the energy of the Sarma phase but its
form is not very enlightening and it will not be needed below.

 The mixed phase is an inhomogeneous phase where a fraction $x$ of the space is in the normal phase with $A$ and $B$ particle densities equal to $\bar n_A$ and $\bar n_B$,  while the remaining  $1-x$  fraction is in the BCS phase with a common density for both species equal to $\bar n$. The densities in each component are  adjusted in such a way that the overall average densities  have  given prescribed values $n_A, n_B$, that is $n_A= x \bar n_A + (1-x) \bar n$ and similarly for the particles $B$. The most favored mixed state for given $n_A, n_B$ is the one with the smallest energy:
  \begin{align}\eqn{mix}
 E_{MIX}(n_A,n_B) =& \text{Min}_{x, \bar n}\ \ \ \left\{(1-x)  \left[ \frac{(6\pi^2 \bar n)^{5/3}}{20\pi^2 M}-\frac{M\Delta^2_0(\bar n) (6\pi^2 \bar n)^{1/3}}{2\pi^2} \right]\right.\nn\\
 &{}+\left. x \frac{(6\pi^2)^{5/3}}{20\pi^2}
 \left[ \frac{1}{M_A}\left( \frac{n_A-(1-x)\bar n}{x}\right)^{5/3} + \frac{1}{M_B}\left(\frac{n_B-(1-x)\bar n}{x}\right)^{5/3} \right]  \right\}.
 \end{align} 
We have disregarded the interface energy between the two components, as those
are small for large enough systems. Also, we assumed that the {\it local}
asymmetries in the densities do not cost additional energy due to long range
forces, as it would be the case if $A$ and $B$ had different
charges~\cite{gapless}. We have to minimize $E_{MIX}$ with respect to normal phase fraction $x$ and the density in the BCS fraction $\bar n$.
 
There are two limiting cases where the comparison between the mixed and Sarma
phases can be done analytically, corresponding to parameters where the
inequalities in \Eq{conditionforsarma} are saturated. 
If $p_B^2-p_A^2=2\sqrt{M_AM_B}\Delta_0$ ($\Delta_s$=0), 
the Sarma state reduces to the normal state. In this case its energy is  
given by
\begin{align}
 E_S &=E_N=\frac{(6\pi^2)^{5/3}}{20\pi^2}  
\left( \frac{n_A^{5/3}}{M_A}+\frac{(n_A+\delta n)^{5/3}}{M_B}   \right)\nn\\
&\cong\frac{(6\pi^2n_A)^{5/3}}{20\pi^2 M}\[1+
+\frac{5 M}{3 M_B}\frac{\delta n}{n_A}
     +\frac{5 M}{9M_B}\left(\frac{\delta n}{n_A}\right)^2+\mathcal{O}(\delta n^3/n_A^3)\],
\end{align}
where $\delta n = n_B-n_A$ is assumed to be small, $\delta n\ll n_A$. 
\begin{figure}[t]
\includegraphics[height=1.9in]{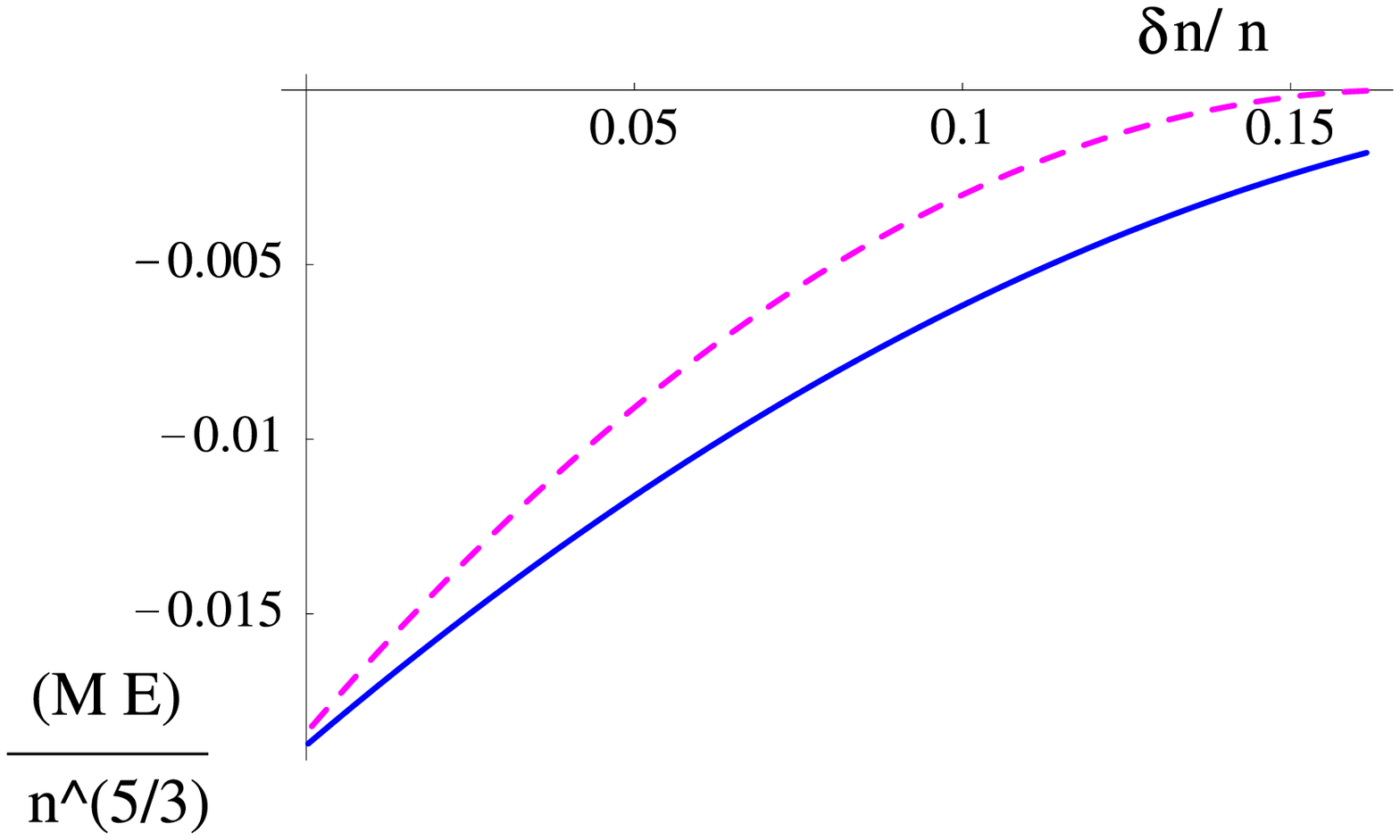}
\hspace{0.2in}
\includegraphics[height=1.7in]{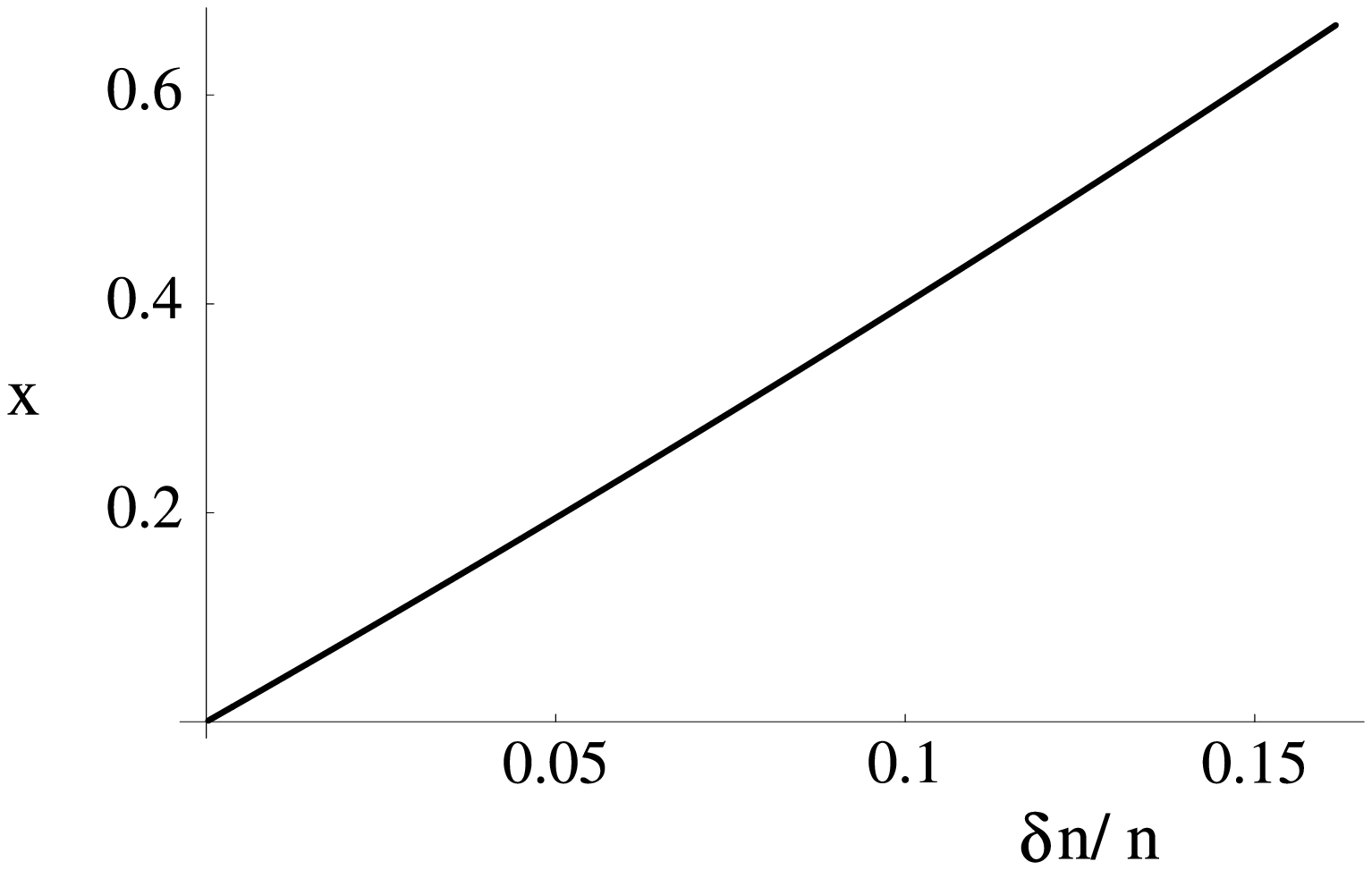}
\caption{\label{mix_and_sarma}\textit{\protect Figure on the left: 
Energy of the Sarma (dashed curve) and
mixed phase (solid curve) relative to the normal phase as a function of $\delta
n/n$ where $n=(n_A+n_B)/2$ with $M_B=7M_A$. 
The values of $n_A$ and $n_B$ were obtained from the Sarma phase for a fixed $a$, 
and  $|a|p_A=0.59$, $0.63\le |a| p_B\le 0.65$.  
Figure on the right: The fraction of normal state 
$x$ as a function of $\delta n/n$ with the same masses and scattering length.}
 } 
\end{figure}

 An upper bound on $E_{MIX}-E_S$ can be obtained by setting the density of the
BCS component of the mixed phase $\bar n=n_A$ and minimizing in relation of
$x$. We have:
 \begin{align}\label{mix-sarma}
 E_{MIX}-E_S \cong -(1-x)(6\pi^2 n_A)^{1/3} \frac{M \Delta_0^2(p_A)}{2\pi^2} 
 + \frac{(6\pi^2 n_A)^{5/3}}{36\pi^2 M_B}\frac{\delta n^2}{n_A^2}\left( \frac{1}{x}-1\right)
 +\mathcal{O}(n_A^{-4/3}\delta n^3),
 \end{align}  whose minimization yields
  \begin{align}
 x = x_{\text{min}} &= \sqrt{\frac{(6\pi^2 n_A)^{4/3}}{18 M M_B \Delta_0^2(p_A)}\frac{\delta n^2}{n_A^2}},\nn\\
 E_{MIX}-E_S &\cong -  (6\pi^2 n_A)^{1/3} \frac{M \Delta_0^2(p_A)}{2\pi^2} (1-x_{\text{min}})^2 < 0.
 \end{align}
Numerical calculations show that the upper bound above is close to the actual minimum. 
 Notice that
 \begin{align}
 \frac{\delta n}{n_A}= \frac{p_B^3-p_A^3}{p_A^3} \cong 3\sqrt{M_AM_B}\frac{\Delta_0}{p_0^2},
 \end{align}
thus, using $\Delta_0(p_A) \cong \Delta_0(p_0)+\mathcal{O}(\Delta^2_0)$, 
we have
  \begin{align}
 x_{\text{min}} \cong \sqrt{\frac{M_A+M_B}{2 M_B}} < 1,
 \end{align} for $M_B> M_A$ as it should be the case. Eq.~(\ref{mix-sarma}) shows that the mixed phase is energetically favored compared to the Sarma phase in one extreme of the window in \Eq{conditionforsarma}.
 
 The other simple limit to analyze corresponds to  $p_B^2-p_A^2=4\sqrt{M_AM_B}\Delta_0$, in which case $\Delta_s=\Delta_0$, $k_1=k_2$,
 $n_A=n_B$ and the Sarma phase reduces to the BCS phase. In this case the mixed phase reduces to the BCS phase too and the energies of both the Sarma and mixed phase are equal to each other. For intermediate values of $p_B^2-p_A^2$ ( still satisfying the constraint in \Eq{conditionforsarma}), the difference $E_{MIX}-E_S$ interpolates between these two extremes, as Fig.~(\ref{mix_and_sarma}) exemplifies. We find that for all reasonable  values of the parameters (that is, where the mean field analysis should apply), and for fixed particle numbers $n_A$ and $n_B$, the mixed phase has a smaller energy than the Sarma phase.
 
 \subsubsection{Fixed total density}

Another interesting situation arises when the total number of particles is
fixed, but conversions between particles of types $A$ and $B$ are allowed. This
is relevant for the physics of high density quark matter where weak interactions
can change the flavor of the quarks. In this situation $n=n_A+n_B$ and
$\delta\mu=(\mu_B-\mu_A)/2$ are fixed 
and the thermodynamic function that should be
minimized is $E-\delta\mu (n_B-n_A)$. 
The non-relativistic formulation presented here is more appropriate for cold
atoms. In cold atom traps, only species with nearly equal masses can 
 convert into one another. 
Further, we consider $\delta\mu=0$ which is relevent for cold atoms for
convenience.  
It is straightforward to see, then, that the
condition for the existence of Sarma state
\Eq{conditionforsarma} is not satisfied. The same also holds when $\delta\mu=0$
and $M_B\gg M_A$.  

The composition in the mixed phase is easy to calculate when $\delta\mu=0$. 
The most favored composition minimizes
\begin{align}
E(n=n_A+n_B)&=\text{Min}_{x, \bar n}\left\{
(1-x)
\left[ \frac{(6\pi^2 \bar n)^{5/3}}{20\pi^2 M}-\frac{M\Delta_0^2(\bar n) (6\pi^2 \bar n)^{1/3}}{2\pi^2} \right]\right.\nn\\
&{}\left.+x\frac{(6\pi^2)^{5/3}}{20\pi^2}\left(m_A^{3/2}
+m_B^{3/2}\right)^{-2/3}\left(\frac{n-2(1-x)\bar n}{x}\right)^{5/3}\right\}.
\end{align}
When $M_A=M_B$, BCS (with $x=0$) is the favored state and when $M_B\gg M_A$
normal phase ($x=1$) with only particles of species $B$ is favored. 
 
 \bigskip

We have considered Fermi gases made up of two species, when an asymmetry on
their densities or chemical potentials tries to push their Fermi surfaces apart,
making pairing more difficult. We find that with either both chemical potentials
or both densities fixed, the most likely ground state is a mixed phase composed
of bubbles of an asymmetric normal state immersed in a sea of the symmetric BCS
phase. It is worth to mention that the LOFF state~\cite{larkin} (in which the condensate varies in space) has lower free energy than the normal and the BCS states. However, the LOFF state can exist only in a very narrow window of asymmetry for the chemical potentials and we ignored this posibility in our discussion.

The space segregation of the excess particles in the mixed phase suggests a possible way of
detecting superfluidity in atomic traps, specially where large gaps are expected
as in the case of ``resonance superfluidity''~\cite{thomas}.  If an optical trap can be filled with an excess number of one of the hyperfine states, the resulting ground state can be imaged in a way that discriminates between them and the bubble structure may become visible. A high concentration of the denser particle species will accumulate  at some point(s) in the trap. The division of the space between the BCS and the normal components is determined by the value of the gap thus, by studying its variation with the variation of the asymmetry, we can infer the existence of superfluidity and even the value of the gap, which is currently an outstanding problem. A better understanding of the surface tension of the interface between the two components is necessary to make this proposal fully quantitative. The qualitative arguments discussed above should be valid even if weak coupling BCS theory is not, as it is the case in the experiments with $^6Li$ ~\cite{thomas}.

\begin{acknowledgments}
We would like to thank Dr. M. Gehm for discussions on the experimental feasibility of observing phase separation in optical traps.
This work was supported by the Director, Office of Energy Research, 
Office of High Energy and Nuclear Physics, and by the Office of 
Basic Energy Sciences, Division of Nuclear Sciences, 
of the U.S. Department of Energy under Contract No.
DE-AC03-76SF00098. The work of one of us (HC) was partially supported by CAPES/Brazil.

\end{acknowledgments}


\end{document}